\pdfoutput=1

\documentclass[11pt,twoside,a4paper,cmspaper,final,collab]{cms-tdr}
\def\svnVersion{d046bae-D}\def\svnDate{2020/07/21}\def\cmsCernNoTag{CERN-EP-2020-019}\def\cmsCernDate{\today}\def\cmsMessage{"Published in the Journal of High Energy Physics as \href{http://dx.doi.org/10.1007/JHEP07(2020)115}{\doi{10.1007/JHEP07(2020)115}.}"}

\begin{document}\cmsNoteHeader{HIN-18-018}

\hyphenation{had-ron-i-za-tion}
\hyphenation{cal-or-i-me-ter}
\hyphenation{de-vices}
\RCS$HeadURL$
\RCS$Id$

\hyphenation{back-ground-sub-traction}
\newcommand{\Deltar}{\ensuremath{\Delta \mathrm{r}}\xspace}
\newcommand{\Njets}{\ensuremath{N_{\text{jets}}}\xspace}

\newlength\cmsTabSkip\setlength{\cmsTabSkip}{1ex}
\providecommand{\NA}{\ensuremath{\text{---}}}

\cmsNoteHeader{HIN-18-018}

\title{Measurement of quark- and gluon-like jet fractions using jet charge in PbPb and $\Pp\Pp$ collisions at 5.02\TeV}

\date{\today}

\abstract{
  The momentum-weighted sum of the electric charges of particles inside a jet, known as jet charge, is sensitive to the electric charge of the particle initiating the parton shower. This paper presents jet charge distributions in $\sqrtsNN=5.02$ TeV lead-lead (PbPb) and proton-proton ($\Pp\Pp$) collisions recorded with the CMS detector at the LHC.  These data correspond to integrated luminosities of 404\mubinv and 27.4\pbinv for PbPb and $\Pp\Pp$ collisions, respectively.  Leveraging the sensitivity of the jet charge to fundamental differences in the electric charges of quarks and gluons, the jet charge distributions from simulated events are used as templates to extract the quark- and gluon-like jet fractions from data. The modification of these jet fractions is examined by comparing $\Pp\Pp$ and PbPb data as a function of the overlap of the colliding Pb nuclei (centrality). This measurement tests the color charge dependence of jet energy loss due to interactions with the quark-gluon plasma. No significant modification between different centrality classes and with respect to $\Pp\Pp$ results is observed in the extracted quark- and gluon-like jet fractions.
}

\hypersetup{
pdfauthor={CMS Collaboration},
pdftitle={Measurement of quark- and gluon-like jet fractions using jet charge in PbPb and pp collisions at 5.02 TeV},
pdfsubject={CMS},
pdfkeywords={CMS, heavy ions, jets}}

\maketitle 

\section{Introduction}

High-momentum partons produced by hard scatterings in heavy ion collisions undergo energy loss as they traverse the quark-gluon plasma (QGP) created in these interactions~\cite{Bjorken:1982tu}. The mechanisms by which these partons lose energy to the medium, as well as their color dependence, are still not fully understood~\cite{Casalderrey-Solana-2007, Majumder-2011}. The particles resulting from the fragmentation and hadronization of these partons can be clustered into jets. Jets are used as parton proxies to examine the properties of the QGP. Parton energy loss manifests itself in various experimental observables including the suppression of high transverse momentum (\pt) hadrons and jets~\cite{Adams:2006yt, Aad:2010bu, Chatrchyan:2011sx, PhysRevLett.91.072301, Adam:2015ewa}, as well as modifications of parton showers~\cite{PhysRevC.98.024908, jetshapes}. These phenomena are collectively referred to as jet quenching~\cite{Bjorken:1982tu}. 

At leading order in quantum chromodynamics, the type of parton that initiates a jet can be distinguished.  The resulting jet can therefore be labeled as a quark, antiquark, or gluon jet. Several recent measurements indicate that the fractions of quark and gluon jets in a sample may be modified as they are expected to suffer different energy loss in the QGP due to their different color charges~\cite{Spousta2016,ALICE_JetSubstructure2019}. This analysis explores the extraction of the fractions of quark and gluon jets from an inclusive jet sample in lead-lead (PbPb) and proton-proton ($\Pp\Pp$) collisions.  This is achieved with a template-fitting method using the ``jet charge`` observable. Jet charge, defined as the momentum-weighted sum of the electric charges of particles inside a jet, is sensitive to the electric charge of the particle initiating a parton shower and can be used to discriminate between gluon- and quark-initiated jets. This observable was initially suggested as a way of measuring the electric charge of a quark~\cite{Feynman:1978} and was first measured in deep inelastic scattering experiments at Fermilab~\cite{berge1980,berge1981}, CERN~\cite{allen1981,allen1982,albanese1984,barlag1982}, and Cornell University~\cite{erickson1979}. 

More recently, jet charge was measured at the LHC in $\Pp\Pp$ collisions by the ATLAS~\cite{Aad_2016} and CMS~\cite{Sirunyan_2017} Collaborations, characterizing the contributions of quark and gluon fragmentation to jet production. At LHC energies, gluon contributions dominate jet production at lower transverse momenta, while the valence quark contributions overtake at higher jet \pt~\cite{Li:2019dre}. According to predictions from the \PYTHIA event generator (version 6.424~\cite{bib_pythia}, tune Z2~\cite{Field}) for $\Pp\Pp$ collisions at 5.02\TeV, gluon jets are expected to constitute about 59\% of a sample of jets with transverse momenta above $120\GeV$. Similarly, up and down (anti)quark jets are predicted to make up about 32\% of the sample with the other 9\% arising from charm, strange and bottom (anti)quark jet contributions. A detailed investigation of jet charge and its applications in heavy ion collisions is motivated by extensive theoretical calculations~\cite{waalewijn2012,heavyions_jetchg_2019,Li:2019dre}. The dependence of the mean and width (standard deviation) of the jet charge distribution on both jet energy and size, can be calculated independently of Monte Carlo (MC) fragmentation models despite the large experimental uncertainty in fragmentation functions~\cite{krohn2013}. This makes jet charge a suitable variable for the determination of quark and gluon jet fractions.

This paper presents the first jet charge measurements in heavy ion collisions along with $\Pp\Pp$ jet charge results at the same center-of-mass energy per nucleon pair($\sqrtsNN$). The analysis uses PbPb and $\Pp\Pp$ data at $\sqrtsNN=5.02$ \TeV, both collected in 2015 with the CMS detector at the CERN LHC. The data correspond to an integrated luminosity of 404$\mubinv$ (27.4$\pbinv$) for PbPb ($\Pp\Pp$) collisions~\cite{CMS-PAS-LUM-16-001}. In heavy ion collisions, the discrimination between jet and background constituents is not straightforward and often impossible on a per-particle basis. In this work, ``background`` is defined as uncorrelated and long-range correlated contributions~\cite{LongRange2011}, as measured at least 1.5 units of relative pseudorapidity ($\Delta\eta$) away from the jet axis~\cite{UICjettrack2015, UICjettrack2016, jetshapes}, which do not arise from the jet-initiating parton shower. Any short-range modifications to either the medium or the jet structure are thus included in the jet ``signal``. The measurements are corrected for detector and background effects using an unfolding procedure, and are presented as a function of the overlap of the colliding Pb nuclei (centrality). The jet charge distributions of light (anti)quark and gluon jets from MC generators are used as templates to fit the inclusive jet charge distribution measured in data. The fractions of quark- and gluon-initiated jets are extracted from this fitting procedure and are referred to as quark- and gluon-like jet fractions.  The results are presented as a function of the minimum \pt threshold of the particles used in the jet charge measurement and also as a function of a \pt weighting factor, $\kappa$~\cite{waalewijn2012}. 

\section{The CMS detector}
\label {sec:Detector}

The central feature of the CMS apparatus is a superconducting solenoid of 6\unit{m} internal diameter, providing a magnetic field of 3.8\unit{T}. Within the solenoid volume are a silicon pixel and strip tracker, a lead tungstate crystal electromagnetic calorimeter (ECAL), and a brass and scintillator hadron calorimeter (HCAL), each composed of barrel and endcap sections.  Two hadron forward (HF) steel and quartz-fiber calorimeters complement the barrel and endcap detectors, extending the calorimeter from the range $\abs{\eta}<3.0$ provided by the barrel and endcap out to $\abs{\eta}<5.2$. Events of interest are selected using a two-tiered trigger system~\cite{Khachatryan:2016bia}. 

In this analysis, jets are reconstructed within the range $\abs{\eta}<1.5$. In the region $\abs{\eta}<1.74$, the HCAL cells have widths of 0.087 in both $\eta$ and azimuth $\phi$. Within the central barrel region corresponding to $\abs{\eta}<1.48$, the HCAL cells map onto $5$$\times$$5$ ECAL crystal arrays to form calorimeter towers projecting radially outwards from the nominal interaction point. Within each tower, the energy deposits in ECAL and HCAL cells are summed to define the calorimeter tower energies, which are subsequently clustered to reconstruct the jet energies and directions~\cite{Chatrchyan:2011ds}. The silicon tracker measures charged-particle tracks within $\abs{\eta}<2.5$. It consists of 1440 silicon pixel and 15\,148 silicon strip detector modules. For charged particles with $1<\pt<10\GeV$ in the barrel region, the track resolutions are typically 1.5\% in \pt and 25--90 (45--150)\mum in the transverse (longitudinal) impact parameter~\cite{TRK-11-001}. A detailed description of the CMS detector, together with a definition of the coordinate system used and the relevant kinematic variables, can be found in Ref.~\cite{Chatrchyan:2008zzk}.

\section{Event selection and simulated event samples}

The PbPb and $\Pp\Pp$ data are selected with a calorimeter-based trigger that uses the anti-\kt jet clustering algorithm with a distance parameter of $R=0.4$~\cite{bib_antikt}. The trigger requires events to contain at least one jet with $\pt>80\GeV$.  This trigger is fully efficient for events containing jets with reconstructed $\pt>100\GeV$. The data selected by this trigger are referred to as ``jet-triggered,`` and corresponds to 3.35 (2.6) million PbPb ($\Pp\Pp$) collision events. Vertex and noise filters are applied to both PbPb and $\Pp\Pp$ data to reduce contamination from noncollision events (\eg, beam-gas interactions), as described in previous analyses~\cite{HIN-15-015, jetshapes}. Additionally, a primary vertex with at least 2 tracks is required to be reconstructed and have a $z$ position (${v_{z}}$) within 15\unit{cm} of the center of the nominal interaction region along the beam axis. In PbPb collisions, the shapes of clusters in the pixel detector are required to be compatible with those expected from a PbPb collision event. The PbPb events are also required to have at least three towers in each of the HF calorimeters with energy deposits of more than $3\GeV$ per tower.

Simulated MC samples are used to evaluate the performance of the event reconstruction, in particular the track reconstruction efficiency and the jet energy response and resolution. The MC samples use the \PYTHIA (version 6.424~\cite{bib_pythia}, tune Z2~\cite{Field}) event generator to describe the hard scattering, parton showering, and hadronization of the partons and are referred to as the \PYTHIA{6} sample. To account for the soft underlying PbPb event, the hard \PYTHIA{6} interactions are embedded into simulated minimum-bias PbPb events produced with \HYDJET (version \textsc{1.383}~\cite{Lokhtin:2005px}).  This minimum-bias event generator is tuned to reproduce global event properties such as the charged-hadron \pt spectrum and particle multiplicity. The combined sample of hard \PYTHIA{6} interactions and soft \HYDJET underlying event is referred to as the \PYTHIA{6}{+}\HYDJET sample. The \GEANTfour~\cite{bib_geant} toolkit is used to simulate the CMS detector response.

The scalar \pt sum of the HF calorimeter towers~($3.0<\abs{\eta}<5.2$) is used to define the event centrality in PbPb events and to divide the event sample into centrality classes, each representing a percentage of the total inelastic hadronic cross section~\cite{Chatrchyan:2012nia}. Events in PbPb collisions are divided into four centrality intervals corresponding to 0--10\% (most central), 10--30\%, 30--50\%, and 50--100\% (most peripheral). 

Because of the large number of nucleon-nucleon interactions in head-on PbPb collisions, jets are more likely to be reconstructed in more central events. Requiring a jet to be present in an event, therefore, biases the data sample toward more central collisions. In comparison, the \PYTHIA{6}{+}\HYDJET sample consists of a flat distribution of jets (from \PYTHIA{6}), as a function of centrality. Thus, a centrality-based reweighting is applied to this MC sample to match the centrality distribution of the jet-triggered PbPb data. An additional reweighting procedure is performed to match the simulated $v_{z}$ distributions to data for both the PbPb and $\Pp\Pp$ samples. The contribution of pile-up in both PbPb and $\Pp\Pp$ collisions is negligible~\cite{CMS-PAS-LUM-16-001}.

\section{Jet and track reconstruction}
\label{sec:jet_track_reco}

The jet reconstruction in PbPb and $\Pp\Pp$ events is performed with the anti-\kt jet algorithm with a distance parameter of $R=0.4$, as implemented in the \FASTJET framework~\cite{fastjet}. Individually calibrated calorimeter towers are used as inputs to the algorithm. Only calorimeter information is used in the jet reconstruction to minimize the bias of the tracking efficiency on the reconstruction of jets. In PbPb collisions, the contributions of the underlying event are subtracted using a two-iteration variant of the ``noise/pedestal subtraction`` technique described in Refs.~\cite{Cacciari:2007fd, Kodolova:2007hd}. In this method, only calorimeter towers outside of the jet area are used in the background estimation after identifying and excluding the jets in the first iteration. The underlying event and pile-up contribution is negligible in $\Pp\Pp$ collisions and therefore do not require any subtraction. In both PbPb and $\Pp\Pp$ events, jet energy is calibrated and the calorimeter response is verified as a function of jet \pt and $\eta$. To account for the variation in detector response with the total number of jet constituents, additional corrections are applied in both collision systems based on the number of charged-particle tracks with $\pt>2\GeV$ within the jet cone (relative angular distance from the jet axis $\Deltar = \sqrt{\smash[b]{(\Delta\eta)^{2}+(\Delta\phi)^{2}}}<0.4$), the jet \pt, and the collision centrality~\cite{jetshapes}. This corrects for a difference in the simulated calorimetric jet energy response between quark and gluon jets and reduces the difference in response between the two jet flavors from 10\% to around 3\%. After reconstruction and offline jet energy calibration, jets are required to have $\pt>120\GeV$ and $\abs{\eta}<1.5$. In this kinematic range, the jet trigger is observed to be fully efficient and the jet energy response and resolution is optimized. Within this selection, it is possible for multiple jets to be selected from the same event. 

For $\Pp\Pp$ collision events, charged-particle tracks are reconstructed using an iterative tracking method~\cite{TRK-11-001} that finds tracks within $\abs{\eta}<2.4$  down to $\pt=0.1\GeV$. For the PbPb data an alternative iterative reconstruction procedure is employed because of the large track multiplicities~\cite{CMS:2012aa,Chatrchyan:2013kwa}. It is capable of reconstructing tracks down to $\pt=0.4\GeV$. The charge of the particle is measured based on the direction of curvature of the reconstructed track. Tracks used in this measurement are required to have a relative \pt uncertainty of less than 10\% (30\%) in PbPb ($\Pp\Pp$) collisions and also satisfy the standard track quality requirements~\cite{HIN-15-015}. For both collision systems, it is required that the significance of the distance of closest approach to at least one primary vertex in the event be less than 3 standard deviations, in order to decrease the likelihood of counting nonprimary charged particles originating from secondary decay products. Tracks with $\pt > 20\GeV$ are required to have an associated energy deposit~\cite{PF_reconstruction2017} of at least half their momentum in the calorimeters to reduce the contribution of misreconstructed tracks with very high $\pt$. In PbPb collisions, tracks must additionally be associated with at least 11 hits and satisfy a fit quality requirement that the $\chi^{2}$, divided by both the number of degrees of freedom and the number of tracker layers hit, be less than 0.15~\cite{HIN-15-015}. The tracking efficiency in $\Pp\Pp$ collisions is approximately 90\% for $\pt>1\GeV$. Track reconstruction is more difficult in the heavy ion environment because of the large track multiplicity, and so the tracking efficiency ranges from approximately 60\% at $\pt=1\GeV$ to about 70\% at $\pt=10\GeV$~\cite{HIN-15-015}.

\section{Jet charge measurement}
\label{sec:jet_chg}

The jet charge is defined as:
\begin{equation} \label{eq:jet_charge_eq}
Q^\kappa = \frac{1}{(\pt^{\text{jet}})^\kappa} \sum_{i\in \text{jet}} {q_i} {p_{\mathrm{T},i}}^\kappa.
\end{equation}
The variable $\pt^{\text{jet}}$ is the transverse momentum of the calorimeter jet. The $q_i$ and $\ensuremath{p_{\mathrm{T},i}}$ symbols refer to the electric charge (in terms of the proton charge $e$) and transverse momentum of the i-th particle in the jet cone, respectively. The $\kappa$ parameter controls the weighting of the jet charge variable to low- and high-\pt particles in the jet cone. Low values of $\kappa$ enhance the contribution from low-\pt particles to the jet charge, and vice versa.

Tracks with $\pt>1\GeV$ that are located within the jet cone ($\Deltar<0.4$) are used in the jet charge measurement. The track \pt threshold of 1\GeV ensures that the MC templates for different flavors used in the fitting procedure are well resolved and also reduces the contributions of uncorrelated and long-range correlated background to the jet charge. Theoretical predictions suggest that a parameter value \pt-weighting factor $\kappa\approx 0.5$ is the most sensitive to the electric charge of the parton initiating the jet in vacuum~\cite{waalewijn2012}. In this analysis, measurements are shown for $\kappa$ values of 0.3, 0.5, and 0.7, and with different selections on the minimum track \pt of 1, 2, 4, and 5\GeV to retain a broad sensitivity to both hard and soft radiation inside jets.  

\section{Corrections for background and detector effects}
\label{sec:unfolding}

To allow for a comparison with future measurements from other experiments or theoretical predictions, the jet charge distributions are unfolded from the detector to the final-state hadron particle level. The jet charge measurements at the detector level are broadened by track reconstruction inefficiencies, and this effect increases with decreasing $\kappa$ values. In PbPb collisions, there is additional smearing that is caused by the background from the underlying event and long-range correlations~\cite{LongRange2011}. The unfolding is performed to account for these effects using the D'Agostini iterative method~\cite{DAgostini1995487,Richardson,Lucy}, as implemented in the \textsc{RooUnfold} software package~\cite{Adye:2011gm}. Response matrices are derived from \PYTHIA{6} and \PYTHIA{6}{+}\HYDJET simulation samples for $\Pp\Pp$ and PbPb collisions, respectively.  

Response matrices in the unfolding procedure are constructed using jet charge distributions measured with reconstructed tracks, and that measured with generator-level particles originating from the hard scattering. To account for the background effects in PbPb collisions, the reconstructed tracks used in constructing the response matrices includes contributions from both the hard scattering and background as modeled by \PYTHIA{6}{+}\HYDJET. As a cross check of the background estimation procedure, a data-driven technique is also used to measure the uncorrelated and long-range correlated contributions, as further discussed in Section~\ref{sec:systematics}. No background correction is required in $\Pp\Pp$ collisions because of the negligible underlying event and pile-up contribution~\cite{CMS-PAS-LUM-16-001}.

The number of iterations in the unfolding procedure trades off bias towards MC with statistical fluctuations. To obtain an optimal number of iterations, reconstructed jet charge distributions from modified samples of \PYTHIA{6} and \PYTHIA{6}{+}\HYDJET are unfolded using the nominal response matrices. Quark and gluon jet fractions are varied by $50\%$ in the modified simulation samples, which is expected to give a good bound on the potential modification of the jet charge distribution in data~\cite{Li:2019dre}. Based on these studies, three to four iterations are used in the unfolding procedure for different selections of threshold track \pt and $\kappa$.

\section{Template fitting}
\label{sec:fitting}

The flavor-tagged jet charge distributions from \PYTHIA{6} at the generator-level are used as templates to fit the unfolded jet charge measurement to estimate the fractions of quark and gluon jets. Measurements from \PYTHIA{6} simulations for jets initiated by up quarks ($\text{mean} = 0.254e$, $\text{width}=0.341e$), down quarks ($\text{mean} = -0.150e$, $\text{width}=0.335e$), and gluons ($\text{mean} = 0.001e$, $\text{width}=0.364e$) are well separated and make up the dominant fractions of the sample. The quoted mean and width values for the different-flavor jets are from measurements at the generator-level, with a minimum track \pt threshold of 1\GeV and a $\kappa$ value of $0.5$. They have statistical uncertainties of less than $0.1\%$. The average jet charge for jets initiated by quarks and gluons varies by less than 1\% as a function of the jet \pt in \PYTHIA{6}, allowing for the stable extraction of the respective jet fractions in the \pt range examined here. In the fitting procedure, the fractions of up antiquark jets ($\cPaqu$) and down antiquark jets ($\cPaqd$), are varied along with the up and down quark jets, respectively. Jets initiated by charm, strange, and bottom (anti)quarks ($\cPaqc$, $\cPqc$, $\cPaqs$, $\cPqs$, $\cPaqb$, and $\cPqb$, respectively) are categorized as ``other flavor" jets and their fractions are fixed during the fitting procedure to reduce the number of degrees of freedom. The fitting procedure takes into consideration the total systematic uncertainty in the jet charge measurements from all sources combined with the statistical uncertainty.

A small fraction of jets have no reconstructed tracks inside the jet cone above the threshold track \pt used in the jet charge measurement. This fraction is negligible for a track \pt threshold of $1\GeV$ in both collision systems but goes up to 10 (6)\% in central PbPb ($\Pp\Pp$) collisions for a track \pt threshold of $5\GeV$. Such jets, with no reconstructed tracks inside the jet cone, are excluded in the fitting procedure, and the fractions of quarks and gluons for such jets are assigned directly from simulation. The related systematic uncertainty is discussed in Section~\ref{sec:systematics}. Previous CMS results have shown a large excess of soft particles in PbPb events relative to $\Pp\Pp$ events up to $\Deltar \sim 1$ from the jet axis, compensated by a relative depletion of higher-$\pt$ tracks~\cite{jetshapes, mpt}. Consequently, the fraction of jets with no high-$\pt$ tracks is observed to be 50\% higher in PbPb data compared to \PYTHIA{6}{+}\HYDJET predictions for the most central collisions.

For a given jet energy, jets with a harder constituent \pt spectrum are more likely to be reconstructed because the calorimeter response does not scale linearly with the incident particle energy, resulting in a bias toward the selection of jets with fewer associated tracks.  On average, quark jets have harder fragmentation than gluon jets and are therefore preferentially reconstructed. Jet energy corrections based on the number of jet constituents are applied to reduce the difference in the response between quark and gluon jets from 10 to around 3\% (see Section~\ref{sec:jet_track_reco}). To compensate for the residual difference, an extra correction factor, based on the deviation from unity in the response, is applied to the extracted fractions of quark- and gluon-like jets~\cite{jetshapes}.

\section{Systematic uncertainties}
\label{sec:systematics}

A number of sources of systematic uncertainty are considered, including effects from the unfolding, tracking efficiencies, background correction, jet reconstruction, and the contributions from ``other flavor" jets. To estimate most systematic uncertainties, a quantity is varied by an appropriate amount in the construction of the response matrix and propagated through the full analysis chain. The fitting procedure is repeated on the varied distributions and the deviation from the nominal results are assigned as systematic uncertainties. The systematic uncertainties from all sources are added in quadrature. The relative uncertainties in the measured jet charge distributions vary for different selections of \pt-weighting factor $\kappa$ and track \pt threshold.

An uncertainty of 5 (4)\% in PbPb ($\Pp\Pp$) data is considered to account for possible differences in track reconstruction between data and simulation, including reconstruction of misreconstructed tracks~\cite{HIN-15-015}. The reconstruction efficiency is varied by this amount when populating the response matrices used in the data unfolding. The resulting jet charge distributions are fit with the generator-level templates and the differences in the extracted fractions of quark- and gluon-like jets, observed to be 1--2\%, are quoted as a source of systematic uncertainty.

From simulation studies, the difference in the tracking efficiencies for positively and negatively charged particles is found to be 0.5\% in both PbPb and $\Pp\Pp$ collisions regardless of the particle \pt. This uncertainty is propagated to the final result in a way similar to what is used for the tracking reconstruction difference between data and simulation, \ie, the unfolding response matrices are modified and any differences after applying the template fitting procedure are taken as systematic uncertainties. 

To study the systematic effect arising from the choice of the MC event generator to produce the response matrix used in the unfolding procedure, a response matrix is formed using a modified \PYTHIA{6} sample with varied quark and gluon jet fractions, and both of these matrices are used to unfold the data. In this study, quark and gluon jet fractions are varied by 50\% from their nominal MC values while populating the modified response matrices. The fitting procedure is repeated on the resulting varied unfolded distributions and the deviation from the nominal fitting results are then assigned as a systematic uncertainty. Other sources of uncertainty in the unfolding procedure include effects from bin-to-bin correlations in the unfolded distribution and the statistical uncertainty in the MC simulation of the response matrix elements. They are propagated using covariance matrices constructed with the \textsc{RooUnfold} software package. These studies result in a relative uncertainty of 4--7\% on the extracted jet fractions. Additional studies are performed using \PYTHIA{8} \textsc{v212}~\cite{Sjostrand:2014zea} tune CUETP8M1~\cite{Khachatryan:2015pea} and \HERWIGpp 2.7.1~\cite{Bahr:2008pv} tune EE5C~\cite{Seymour:2013qka} event generators, neither of which are observed to describe the jet spectra in $\Pp\Pp$ data very well. After reweighting the jet spectra in these MC samples to match data, while jet charge distributions from \PYTHIA{8} are in very good agreement with those from \PYTHIA{6} and data, \HERWIGpp overestimates the width of the data jet charge distributions and are hence not used in systematic uncertainty studies.

The systematic uncertainty due to the jet energy resolution is estimated by changing the jet energy resolution by 5\% to cover the corresponding uncertainty~\cite{Chatrchyan:2011ds}, followed by a comparison of the modified spectra with the nominal spectrum. The corresponding differences in the extracted quark- and gluon-like jet fractions, estimated from repeating the fitting procedure on the smeared jet charge distributions, are 1--3\% and are included as systematic uncertainties. The effects of the angular resolution of the jet axis are negligible in the jet charge measurements.

To study the background modelling uncertainty in PbPb collisions, the response matrices are also built using a data-driven reference event technique to estimate the uncorrelated and long-range correlated background contributions. The jet charge is measured using jets in a jet-triggered ``signal`` event and tracks from a separate minimum-bias ``reference`` event which is required to have a $v_{z}$ within 1\unit{cm} and collision centrality within 2.5\% of the signal event. The signal and reference events must also have similar charged particle multiplicities outside of the jet cone. The background obtained from the reference event technique is observed to be in close agreement with that for \HYDJET and the resulting uncertainty is less than $1\%$. No background subtraction is performed in $\Pp\Pp$ due to its negligible effect, and hence no corresponding systematic uncertainty is assigned.

The contribution from jets with no tracks in the jet cone above a \pt threshold, which are excluded in the fitting procedure, to the gluon-like jet fraction measurements is assigned from MC. The difference in the fraction of such jets between data and MC increases with increasing track \pt threshold and with more central collisions because of the observed depletion of high-$\pt$ tracks in PbPb collisions~\cite{jetshapes, mpt}. This difference is less than 1\% in $\Pp\Pp$ collisions but can reach $4.5\%$ in PbPb collisions. It is assigned as a systematic uncertainty.   

In PbPb data, there is an additional bias toward selecting jets that are reconstructed on upward fluctuations in the underlying event. Since the jet spectrum is steeply falling, more jets on upward fluctuations are included in the sample than jets on downward fluctuations are excluded resulting in an uncertainty of up to $10\%$ in the measured particle multiplicity in central PbPb events~\cite{jetshapes}. This effect is observed to be included in the reconstructed jet charge measurements in simulation as well, so the difference in this bias between data and MC is used to calculate the corresponding systematic uncertainty. To calculate this difference, distributions of the particle multiplicities within cones ($\Deltar<0.4$), chosen randomly in detector $\eta$ and $\phi$, are compared between minimum-bias data and MC events~\cite{jetshapes, mpt} and are found to be in very good agreement with each other. The difference is propagated through the analysis chain and the resulting deviation from the nominal results are observed to be negligible.

To assess the effects of the statistical uncertainties from the MC templates on the final results, 1000 pseudo-experiments are performed by generating smeared jet charge templates based on its statistical uncertainty and repeatedly fitting the data measurements using these templates. The distributions of extracted gluon-like jet fractions from the pseudo-experiment fits have a variance of 3\% or less, which is assigned as a systematic uncertainty due to limited MC event count.

Finally, the effect of fixing the ``other flavor" jet fractions in the fitting procedure is analyzed. The ``other flavor" jets, comprising c ($2.9\%$), s ($4.7\%$) and b ($1.7\%$) (anti)quarks, are each varied by their total amount in the fitting procedure and the resulting deviation from the nominal fitting result is propagated as a systematic uncertainty. 

A summary of the range of systematic uncertainties for results is shown in Table~\ref{tab:sys_ptcut} for different selections of $\kappa$ and track \pt threshold values.

\begin{table}[htbp]
\centering
\topcaption{Relative systematic uncertainties in percentage for the measurements of gluon-like jet fractions in $\Pp\Pp$ and PbPb events. The PbPb results are given in intervals of centrality. When an uncertainty range is given, the range of the values are the maximum variation in the fractions for different selections on $\kappa$ and track \pt threshold values.}
\begin{tabular}{cccccc}
\hline
& $\Pp\Pp$ & \multicolumn{4}{c}{PbPb centrality intervals} \\
Source                                   &  &        50--100\% &   30--50\% &   10--30\%&    0--10\%  \\
\hline
Response matrix modelling                & 4--6&     5--7.5&       5--7.5&      5--7.5&      5--7.5 \\
Monte Carlo event count                  & 1.5 &     3 &           3 &          3 &          3  \\ 
Jet energy resolution                    & 1--1.5&   2 &           2 &          2--3 &       2--3  \\
Tracking efficiency (data/simulation)    & 1&        2 &           2 &          2 &          2    \\
Tracking efficiency (positive/negative)  & 0.5--1&   1--1.5&       1--1.5 &     0.5--1.5 &   0.5--1.5 \\
Jets with no tracks                      & 0.1&      0.2--1&       0.4--2 &     0.4--3 &     0.5--4.5 \\
Unfolding procedure                      & 0.5&      0.7 &         0.8&         1.1&         1.4 \\ 
Background modelling and fluctuation     & \NA &     0.5 &         0.5 &        1 &          1 \\ 
``Other flavor" jets                     & 1 &       1 &           1 &          1 &          1 \\
Total                                    & 4--5 &    7--8 &        7--8 &       7--8 &       7--9  \\
\hline
\end{tabular}
\label{tab:sys_ptcut}
\end{table}

\section{Results}
\label{sec:results}

\begin{figure}[t]
\centering
\includegraphics[width=\textwidth]{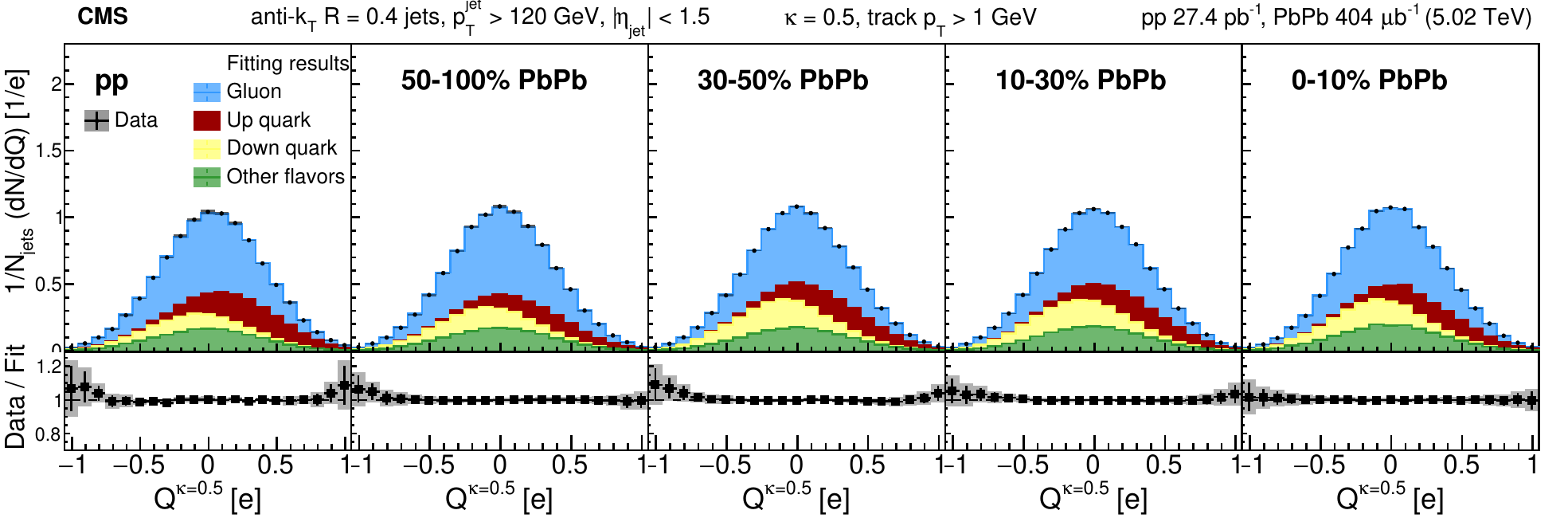}
\caption{(Upper) Unfolded jet charge measurements shown for inclusive jets in data along with the extracted fractions of up, and down quark jets, gluon jets, and the ``other flavor" jets.  The systematic and statistical uncertainties in the distributions are shown by the shaded regions and vertical bars, respectively. The jet charge measurements shown here are for the $\pt$-weighting factor $\kappa = 0.5$ and a minimum track \pt of 1\GeV. (Lower) Ratio of the jet charge measurements to the results of template fits.}
\label{fig:fitting_ratio}
\end{figure}

The unfolded jet charge measurements, normalized to the total number of jets in the sample (\Njets), are shown in the upper panels of Fig.~\ref{fig:fitting_ratio} with solid black points for a sample selection with a minimum track \pt of 1\GeV and \pt-weighting factor $\kappa=0.5$. The results are shown for $\Pp\Pp$ and different event centrality bins in PbPb. The extracted fraction of quark- and gluon-initiated jets is displayed as a set of stacked histograms. Figure~\ref{fig:fitting_ratio} also shows the ratio of the data over the template fit results in the lower panels, and no significant deviation from unity is observed in the entire fitting range. The fractions of up and down quarks are observed to be significantly different between pp and PbPb collisions, as expected from an enhanced contribution of valence down quarks in lead collisions (having 126 neutrons and 82 protons in each nucleus). The jet charge measurements and fit results for other minimum track \pt and $\kappa$ selections are shown in figures~\ref{fig:jetchg_dist_fit_data_trk}-\ref{fig:jetchg_dist_fit_data_kappa} in the appendix.   

\begin{figure}[th]
\centering
\includegraphics[width=\textwidth]{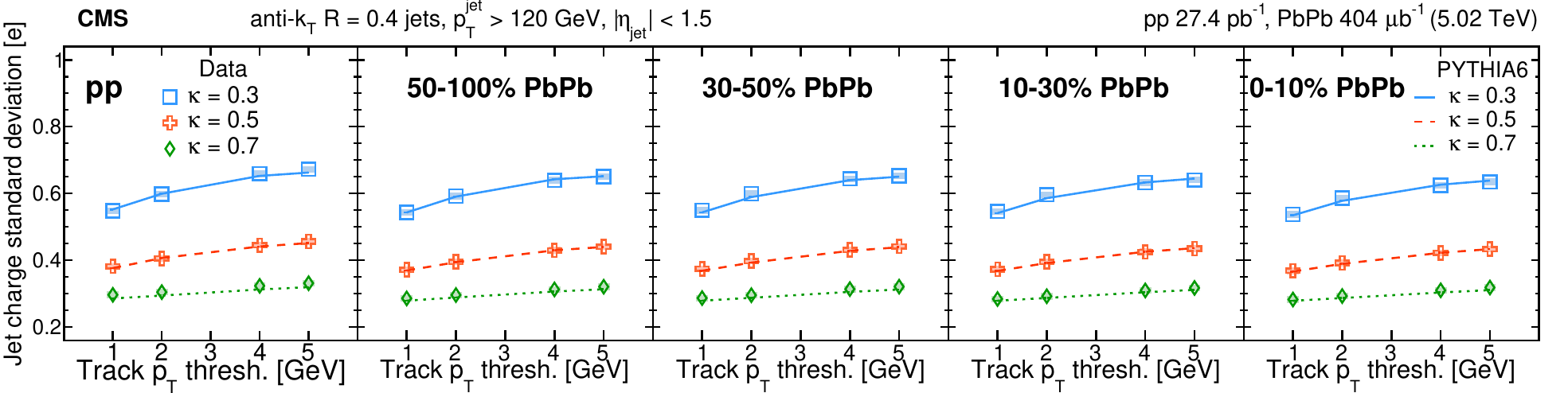}
\caption{The standard deviation of the jet charge distributions with different track \pt thresholds and $\kappa$ values for $\Pp\Pp$ collisions and in the various event centrality bins for PbPb collisions compared with the \PYTHIA{6} prediction. The systematic and statistical uncertainties in the standard deviation measurements are shown by the shaded regions and vertical bars, respectively.}
\label{fig:jetcharge_width}
\end{figure}

The widths (standard deviations) of the unfolded data jet charge distributions in different PbPb event centrality bins and in $\Pp\Pp$, with various track \pt thresholds and $\kappa$ values, are shown in Fig.~\ref{fig:jetcharge_width}. They are also compared to generator-level predictions from \PYTHIA{6} with matching track \pt and $\kappa$ selections in Fig.~\ref{fig:jetcharge_width}. The data (simulation) results for $\kappa = 0.3$, 0.5, and 0.7, are shown by the blue squares (solid lines), red crosses (dashed lines), and green diamonds (dotted lines), respectively. The measured standard deviations tend to increase as a function of the minimum track \pt and decrease with increasing $\kappa$ value. Theoretical predictions incorporating color-charge dependence into jet energy loss calculations predict that stronger quenching of gluon jets will result in a reduced fraction of gluon-initiated jets in the observed jet sample from PbPb collisions compared to that in $\Pp\Pp$ data~\cite{heavyions_jetchg_2019}. The mean of the jet charge distribution for gluon-initiated jets is consistently predicted to be zero in various MC simulations, while that of quark jets is nonzero. A decrease in the fraction of gluons in a quenched jet sample would hence lead to an effective increase in the standard deviation of the measured jet charge distribution. Figure~\ref{fig:jetcharge_width} summarizes standard deviations measured for all track \pt selections and $\kappa$ values studied. The generator-level \PYTHIA{6} predictions agree with the measured widths for $\Pp\Pp$ events. No strong modifications are observed in the widths of the jet charge distributions in central PbPb collisions compared to the peripheral events. While the PbPb width results cannot be directly compared to the $\Pp\Pp$ reference due to different up and down quark contents in protons and Pb nuclei, generator-level \PYTHIA{6} predictions adjusted for this difference reproduce the observed widths of jet charge measurements for all PbPb collision centralities.

\begin{figure}[thb]
\centering
\includegraphics[width=\textwidth]{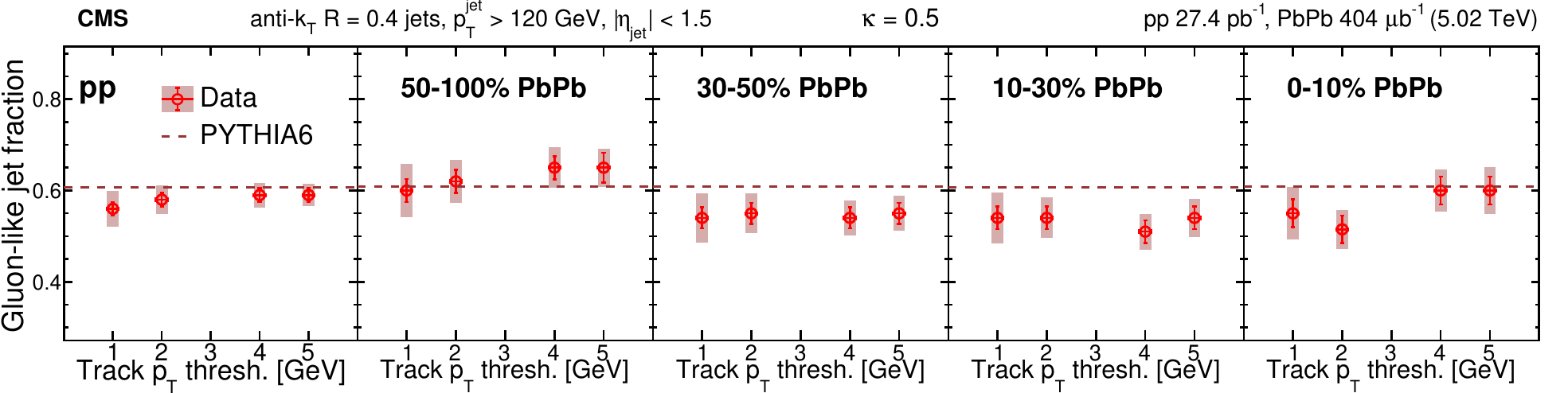}
\caption{Fitting results for the extraction of gluon-like jet fractions in $\Pp\Pp$ and PbPb data shown for different track \pt threshold values and event centrality bins in PbPb collisions. The systematic and statistical uncertainties are represented by the shaded regions and vertical bars, respectively. The predictions for the gluon jet fractions from \PYTHIA{6} are shown in dashed red lines.}
\label{fig:fitting_qg_ptcut}
\end{figure}

\begin{figure}[h]
\centering
\includegraphics[width=\textwidth]{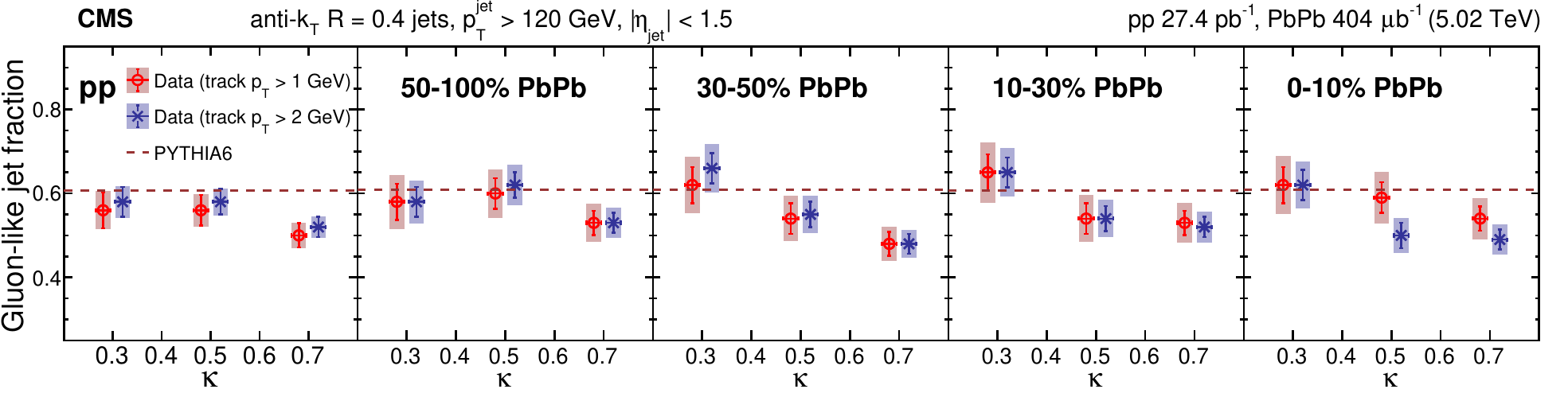}
\caption{Fitting results for the extraction of gluon-like jet fractions in $\Pp\Pp$ and PbPb data shown for \pt-weighting factor $\kappa$ values of 0.3, 0.5, and 0.7 in different event centrality bins in PbPb. The markers for track $\pt > 1$ and $>$2\GeV have been separated horizontally for clarity. The systematic and statistical uncertainties are represented by the shaded regions and vertical bars, respectively. The predictions for the gluon jet fractions from \PYTHIA{6} are shown in dashed red lines.}
\label{fig:fitting_qg_kappa}
\end{figure}

The results for the quark- and gluon-like jet fractions in an inclusive sample are shown in Fig.~\ref{fig:fitting_qg_ptcut} as a function of the track \pt threshold.  Figure~\ref{fig:fitting_qg_kappa} shows the same quantities as a function of $\kappa$ for track $\pt > 1$ and $>$2\GeV, with red circles and blue crosses, respectively. The systematic uncertainties are shown by the shaded regions while the statistical uncertainties, combined with the fit uncertainties, are shown by the solid vertical bars. Only the gluon jet fitting results are shown in Figs.~\ref{fig:fitting_qg_ptcut} and~\ref{fig:fitting_qg_kappa} for clarity but it should be inferred that the quark jets make up the rest of the inclusive sample. 

Previous CMS measurements have shown a strong modification in the distribution of low-\pt tracks relative to the jet axis in PbPb collisions with respect to $\Pp\Pp$ collisions~\cite{jetshapes, mpt}. In-medium gluon radiation and a wake-like response of the QGP to the propagating parton are two of the proposed explanations for this modification~\cite{Casalderrey-Solana:2016}, neither of which are expected to modify the jet charge considerably. From Figs.~\ref{fig:fitting_qg_ptcut} and~\ref{fig:fitting_qg_kappa}, no significant modification is observed in the relative fractions of the quark- and gluon-like jets in central PbPb collisions compared to peripheral PbPb and $\Pp\Pp$ collisions. The relative jet fractions are also observed to be unmodified when calculated using a range of different track \pt thresholds or $\kappa$ values.

\section{Summary}
\label{sec:Summary}

Jet charge, defined as the momentum-weighted sum of the electric charges of particles inside a jet, is measured for the first time in heavy ion collisions and is presented along with $\Pp\Pp$ results at the same energy. The analysis uses lead-lead (PbPb) and proton-proton ($\Pp\Pp$) collision data collected with the CMS detector at a nucleon-nucleon center-of-mass energy of 5.02\TeV. The unfolded jet charge distributions, measured using the jet constituents with transverse momentum $\pt > 1\GeV$ for jets having $\pt > 120\GeV$ and pseudorapidity $\abs{\eta} < 1.5$, are presented. The widths of the jet charge distributions for $\Pp\Pp$ collisions are in good agreement with predictions from the event generator \PYTHIA{6} and are shown to be independent of PbPb collision centrality. The jet charge distributions for quark- and gluon-initiated jets from \PYTHIA{6} events are used as fitting templates to estimate the respective contributions in the measured jet samples. The gluon-like jet fractions extracted from these template fits are found to be similar between $\Pp\Pp$ data and all studied PbPb centrality ranges. These are the first measurements in heavy ion collisions which exploit the electric charge of the initiating parton to discriminate between quark and gluon jets. No evidence is seen for a significant decrease (increase) in gluon-like (quark-like) prevalence in a sample of jets with $\pt > 120\GeV$ in PbPb collisions. These observations do not support recent interpretations of other heavy ion results~\cite{Spousta2016,ALICE_JetSubstructure2019}, which are based on a decreased (increased) gluon (quark) fraction caused by color-charge dependent jet quenching.

\begin{acknowledgments}
  We congratulate our colleagues in the CERN accelerator departments for the excellent performance of the LHC and thank the technical and administrative staffs at CERN and at other CMS institutes for their contributions to the success of the CMS effort. In addition, we gratefully acknowledge the computing centers and personnel of the Worldwide LHC Computing Grid for delivering so effectively the computing infrastructure essential to our analyses. Finally, we acknowledge the enduring support for the construction and operation of the LHC and the CMS detector provided by the following funding agencies: BMBWF and FWF (Austria); FNRS and FWO (Belgium); CNPq, CAPES, FAPERJ, FAPERGS, and FAPESP (Brazil); MES (Bulgaria); CERN; CAS, MoST, and NSFC (China); COLCIENCIAS (Colombia); MSES and CSF (Croatia); RPF (Cyprus); SENESCYT (Ecuador); MoER, ERC IUT, PUT and ERDF (Estonia); Academy of Finland, MEC, and HIP (Finland); CEA and CNRS/IN2P3 (France); BMBF, DFG, and HGF (Germany); GSRT (Greece); NKFIA (Hungary); DAE and DST (India); IPM (Iran); SFI (Ireland); INFN (Italy); MSIP and NRF (Republic of Korea); MES (Latvia); LAS (Lithuania); MOE and UM (Malaysia); BUAP, CINVESTAV, CONACYT, LNS, SEP, and UASLP-FAI (Mexico); MOS (Montenegro); MBIE (New Zealand); PAEC (Pakistan); MSHE and NSC (Poland); FCT (Portugal); JINR (Dubna); MON, RosAtom, RAS, RFBR, and NRC KI (Russia); MESTD (Serbia); SEIDI, CPAN, PCTI, and FEDER (Spain); MOSTR (Sri Lanka); Swiss Funding Agencies (Switzerland); MST (Taipei); ThEPCenter, IPST, STAR, and NSTDA (Thailand); TUBITAK and TAEK (Turkey); NASU (Ukraine); STFC (United Kingdom); DOE and NSF (USA). 
 
  \hyphenation{Rachada-pisek} Individuals have received support from the Marie-Curie program and the European Research Council and Horizon 2020 Grant, contract Nos.\ 675440, 752730, and 765710 (European Union); the Leventis Foundation; the A.P.\ Sloan Foundation; the Alexander von Humboldt Foundation; the Belgian Federal Science Policy Office; the Fonds pour la Formation \`a la Recherche dans l'Industrie et dans l'Agriculture (FRIA-Belgium); the Agentschap voor Innovatie door Wetenschap en Technologie (IWT-Belgium); the F.R.S.-FNRS and FWO (Belgium) under the ``Excellence of Science -- EOS" -- be.h project n.\ 30820817; the Beijing Municipal Science \& Technology Commission, No. Z191100007219010; the Ministry of Education, Youth and Sports (MEYS) of the Czech Republic; the Deutsche Forschungsgemeinschaft (DFG) under Germany's Excellence Strategy -- EXC 2121 ``Quantum Universe" -- 390833306; the Lend\"ulet (``Momentum") Program and the J\'anos Bolyai Research Scholarship of the Hungarian Academy of Sciences, the New National Excellence Program \'UNKP, the NKFIA research grants 123842, 123959, 124845, 124850, 125105, 128713, 128786, and 129058 (Hungary); the Council of Science and Industrial Research, India; the HOMING PLUS program of the Foundation for Polish Science, cofinanced from European Union, Regional Development Fund, the Mobility Plus program of the Ministry of Science and Higher Education, the National Science Center (Poland), contracts Harmonia 2014/14/M/ST2/00428, Opus 2014/13/B/ST2/02543, 2014/15/B/ST2/03998, and 2015/19/B/ST2/02861, Sonata-bis 2012/07/E/ST2/01406; the National Priorities Research Program by Qatar National Research Fund; the Ministry of Science and Education, grant no. 14.W03.31.0026 (Russia); the Tomsk Polytechnic University Competitiveness Enhancement Program and ``Nauka" Project FSWW-2020-0008 (Russia); the Programa Estatal de Fomento de la Investigaci{\'o}n Cient{\'i}fica y T{\'e}cnica de Excelencia Mar\'{\i}a de Maeztu, grant MDM-2015-0509 and the Programa Severo Ochoa del Principado de Asturias; the Thalis and Aristeia programs cofinanced by EU-ESF and the Greek NSRF; the Rachadapisek Sompot Fund for Postdoctoral Fellowship, Chulalongkorn University and the Chulalongkorn Academic into Its 2nd Century Project Advancement Project (Thailand); the Kavli Foundation; the Nvidia Corporation; the SuperMicro Corporation; the Welch Foundation, contract C-1845; and the Weston Havens Foundation (USA). 
\end{acknowledgments}

\bibliography{auto_generated}

\clearpage

\appendix

\section{Jet charge measurements} 
\label{app:jetchg_fits_data}

\begin{figure}[h!]
\centering
\includegraphics[width=1.\textwidth]{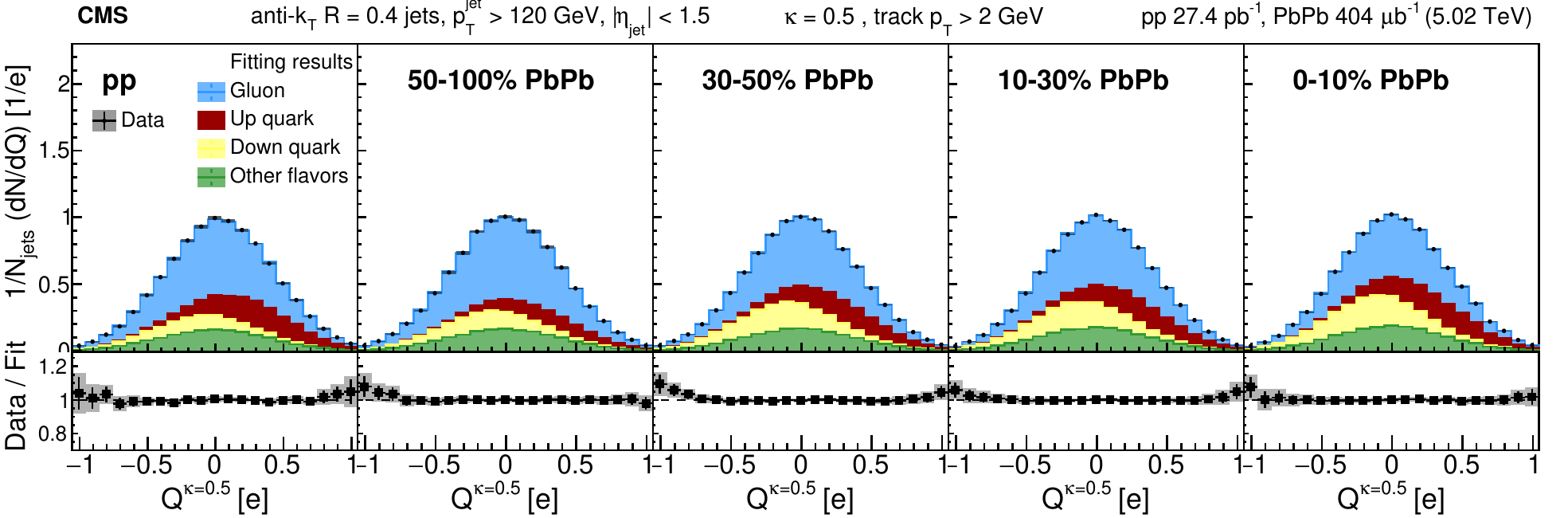}\\
\vspace{\baselineskip}
\includegraphics[width=1.\textwidth]{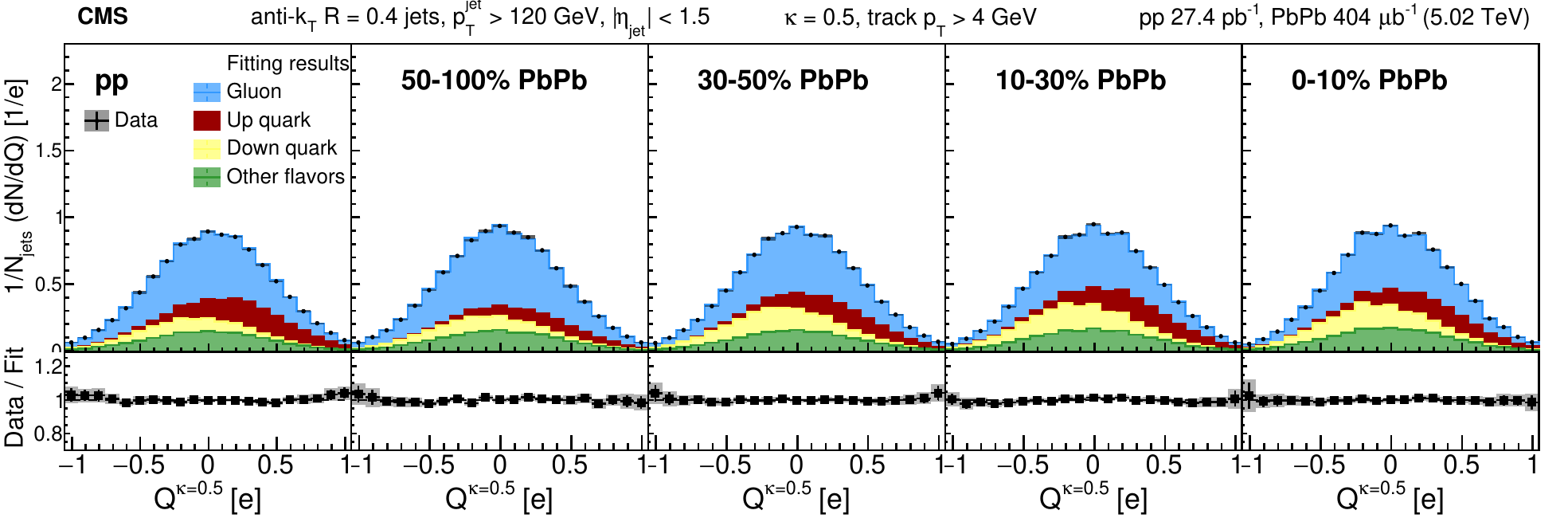}\\
\vspace{\baselineskip}
\includegraphics[width=1.\textwidth]{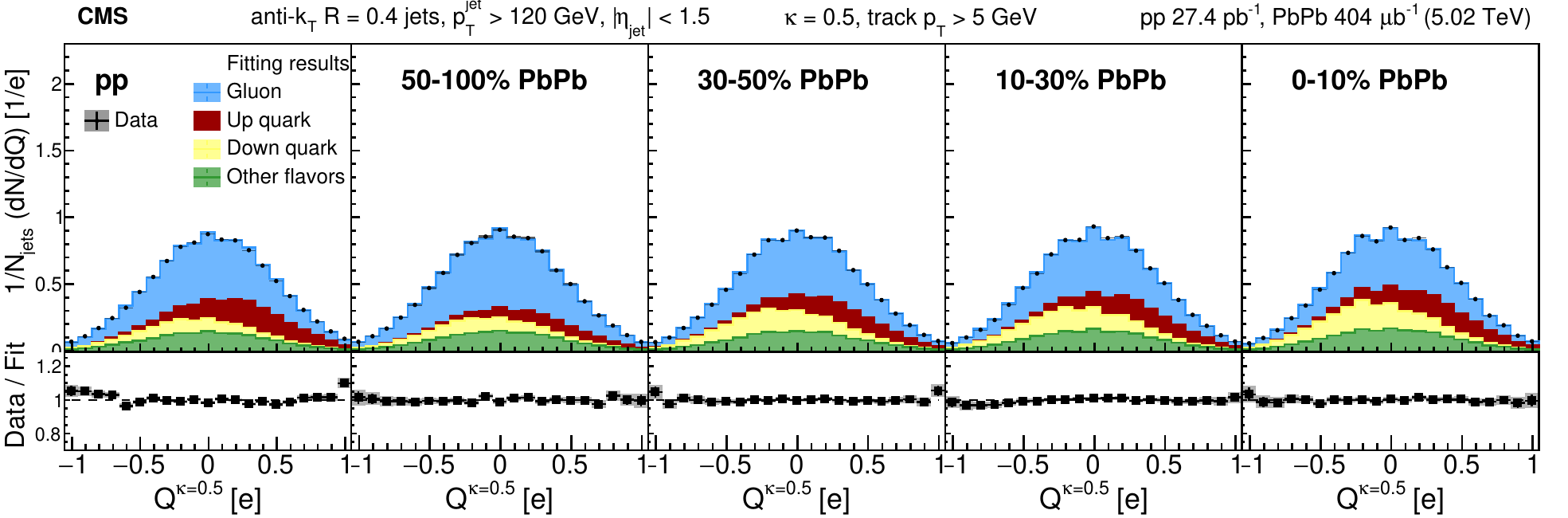}
\caption{(Upper row of each figure) Unfolded jet charge measurements shown for inclusive jets in data along with the extracted fractions of up, and down quark jets, gluon jets, and the ``other flavor" jets.  The systematic and statistical uncertainties in the distributions are shown by the shaded regions and vertical bars, respectively. The jet charge measurements shown here are for $\kappa = 0.5$ and a minimum track \pt of 2, 4, and 5\GeV (top, middle, and bottom, respectively). (Lower row of each figure) Ratio of the jet charge measurements to the results of template fits.}
\label{fig:jetchg_dist_fit_data_trk}

\end{figure}

\clearpage

\begin{figure}[h!]
\centering
\includegraphics[width=1.\textwidth]{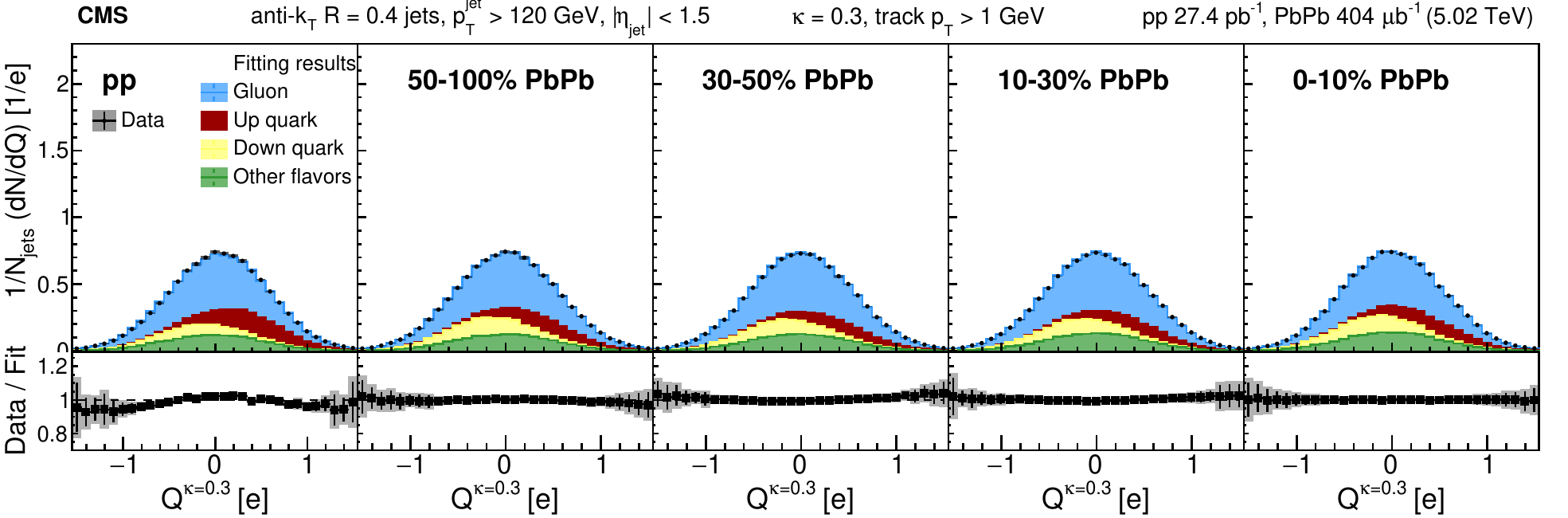}\\
\vspace{\baselineskip}
\includegraphics[width=1.\textwidth]{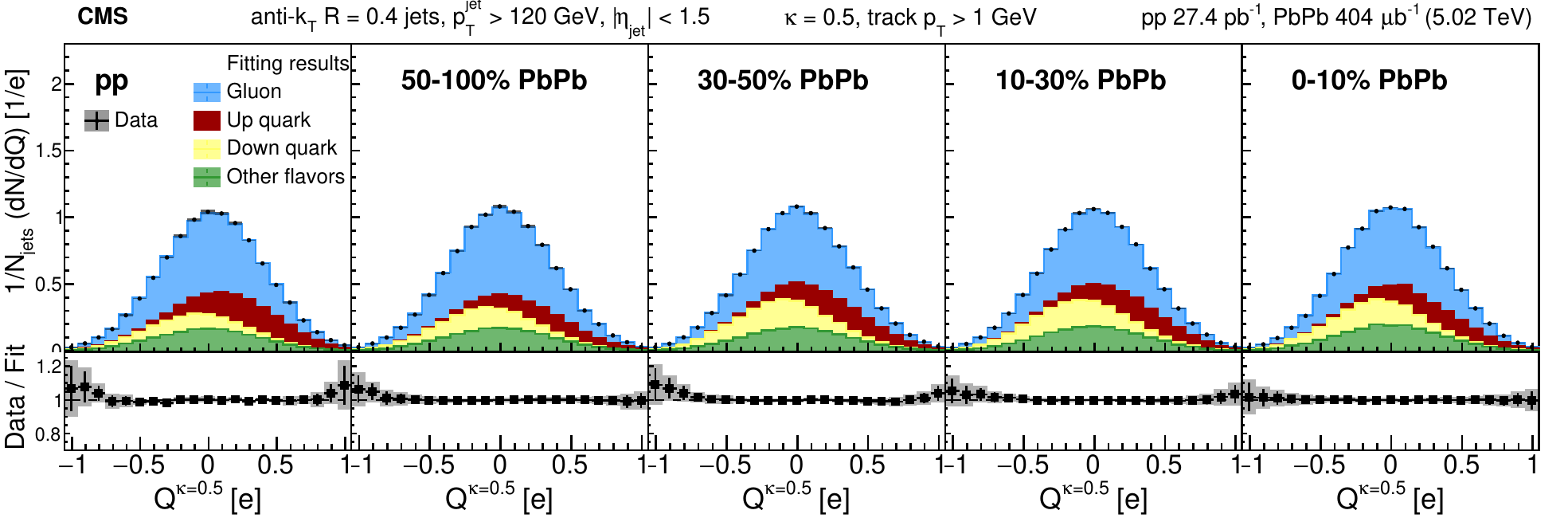}\\
\vspace{\baselineskip}
\includegraphics[width=1.\textwidth]{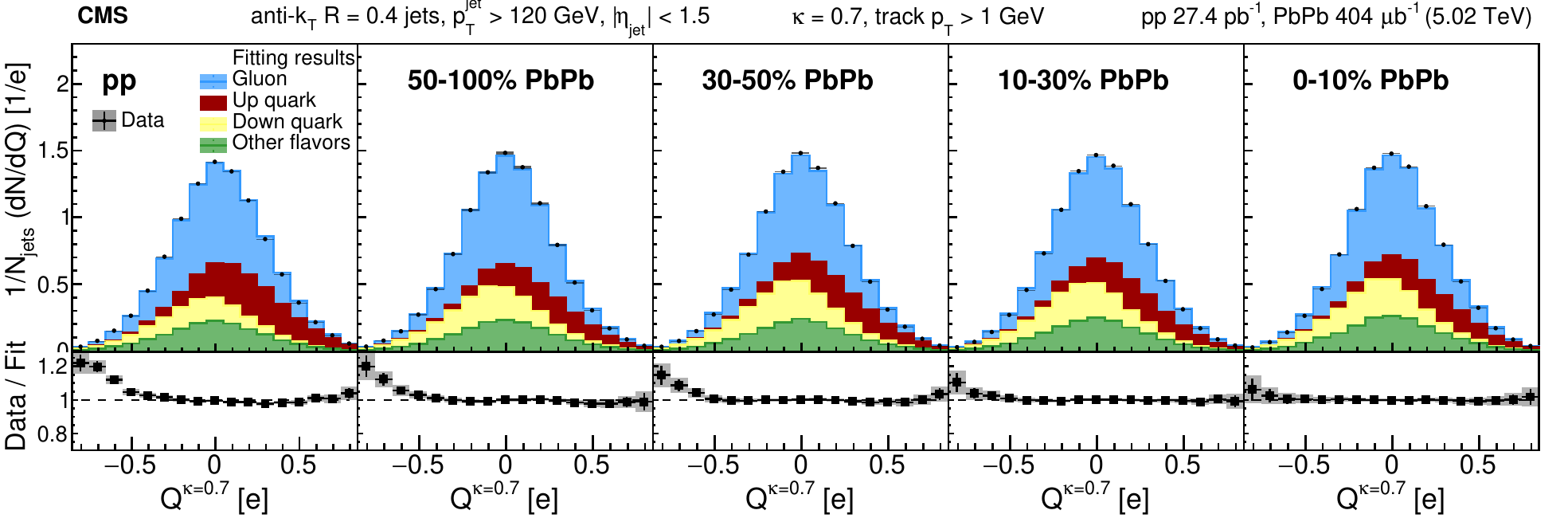}
\caption{(Upper row of each figure) Unfolded jet charge measurements shown for inclusive jets in data along with the extracted fractions of up, and down quark jets, gluon jets, and the ``other flavor" jets.  The systematic and statistical uncertainties in the distributions are shown by the shaded regions and vertical bars, respectively. The jet charge measurements shown here are for a minimum track \pt of 1\GeV and a $\kappa$ value of 0.3, 0.5, and 0.7 (top, middle, and bottom, respectively). (Lower row of each figure) Ratio of the jet charge measurements to the results of template fits.}
\label{fig:jetchg_dist_fit_data_kappa_trkpt1}

\end{figure}

\clearpage

\begin{figure}[h!]
\centering
\includegraphics[width=1.\textwidth]{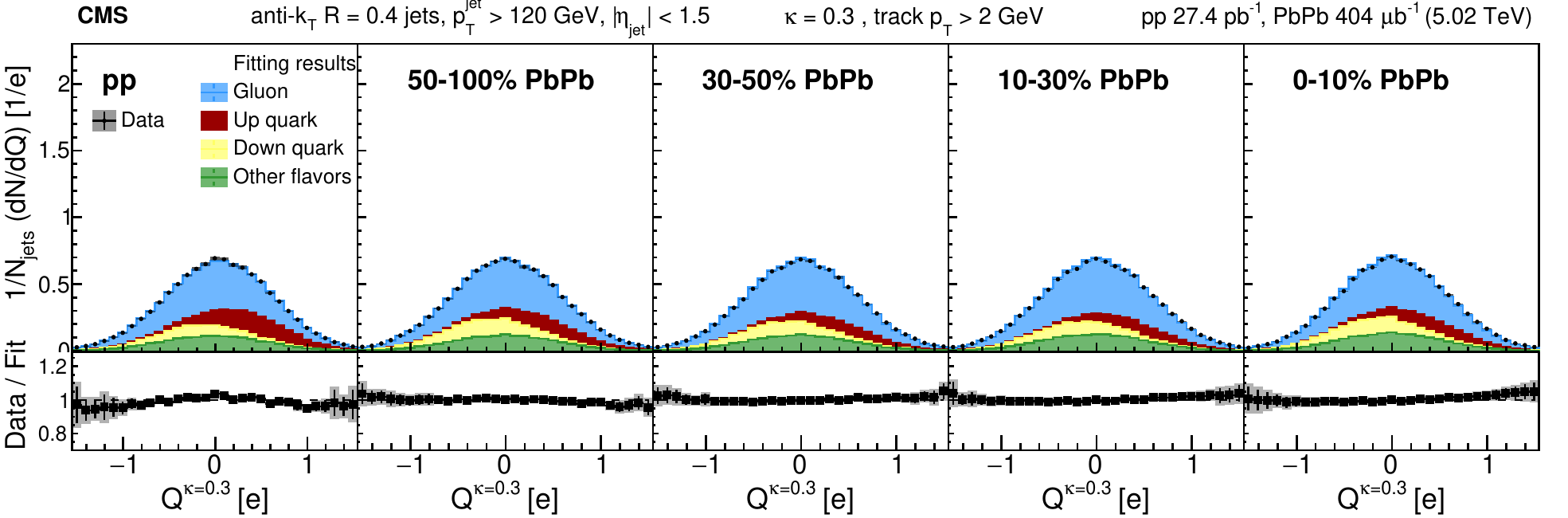}\\
\vspace{\baselineskip}
\includegraphics[width=1.\textwidth]{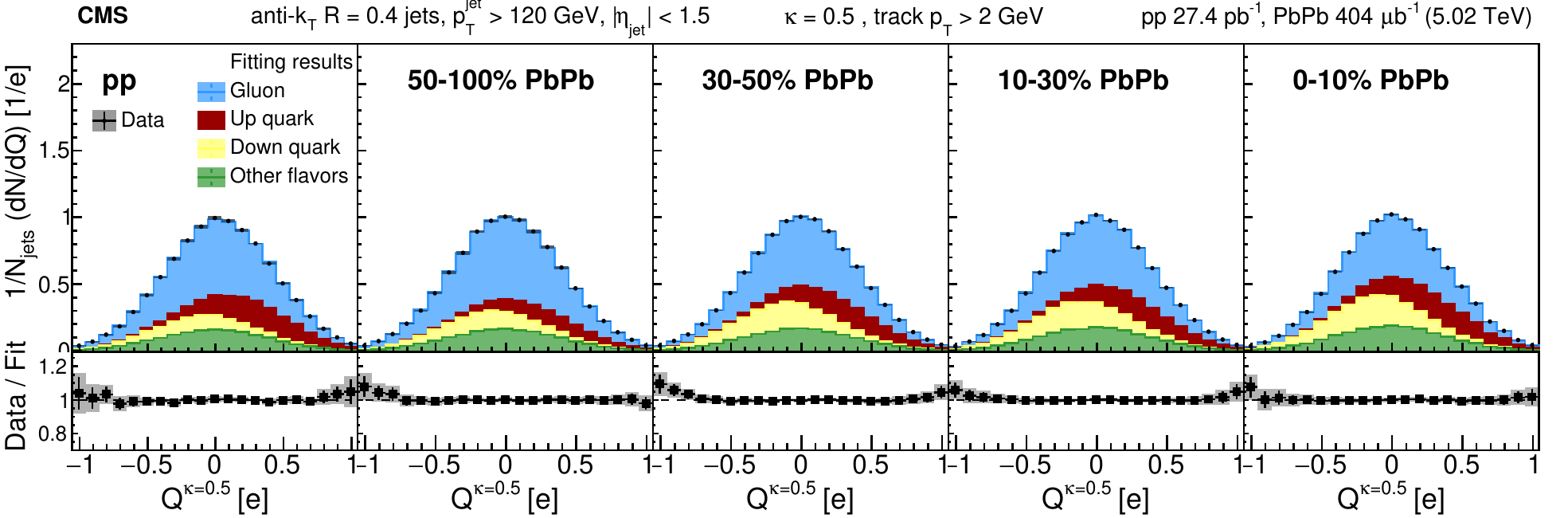}\\
\vspace{\baselineskip}
\includegraphics[width=1.\textwidth]{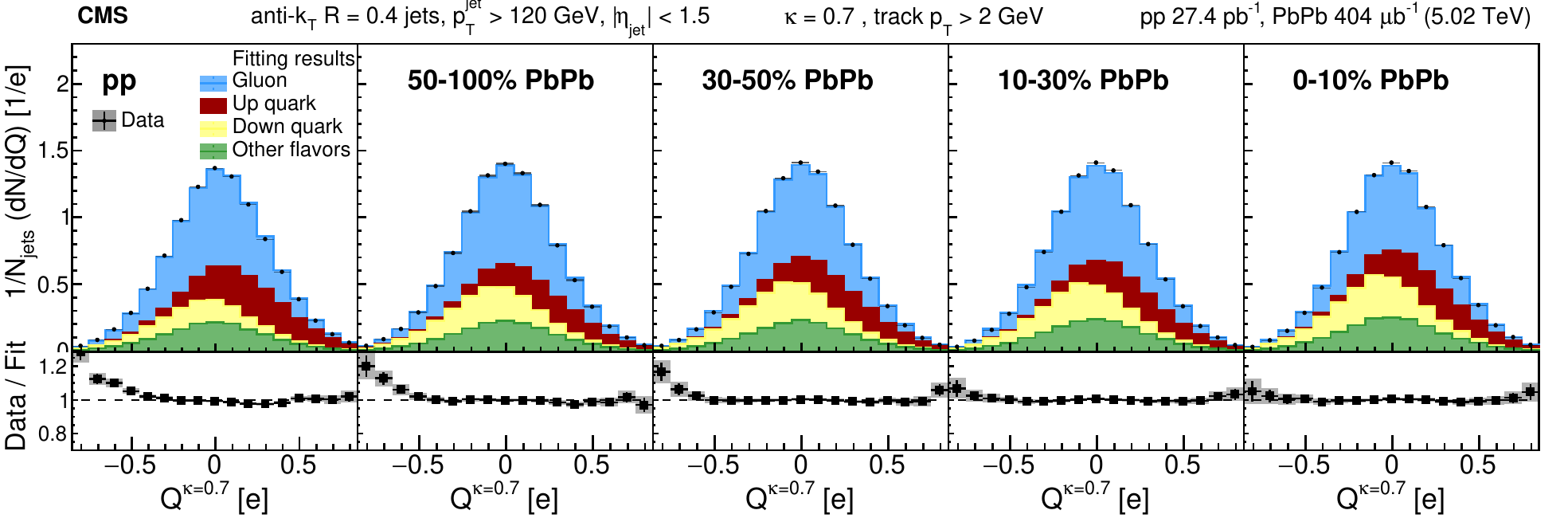}
\caption{
(Upper row of each figure) Unfolded jet charge measurements shown for inclusive jets in data along with the extracted fractions of up, and down quark jets, gluon jets, and the ``other flavor" jets.  The systematic and statistical uncertainties in the distributions are shown by the shaded regions and vertical bars, respectively. The jet charge measurements shown here are for a minimum track \pt of 2\GeV and a $\kappa$ value of 0.3, 0.5, and 0.7 (top, middle, and bottom, respectively). (Lower row of each figure) Ratio of the jet charge measurements to the results of template fits.}
\label{fig:jetchg_dist_fit_data_kappa}

\end{figure}

\clearpage
\cleardoublepage \section{The CMS Collaboration \label{app:collab}}\begin{sloppypar}\hyphenpenalty=5000\widowpenalty=500\clubpenalty=5000\vskip\cmsinstskip
\textbf{Yerevan Physics Institute, Yerevan, Armenia}\\*[0pt]
A.M.~Sirunyan$^{\textrm{\dag}}$, A.~Tumasyan
\vskip\cmsinstskip
\textbf{Institut f\"{u}r Hochenergiephysik, Wien, Austria}\\*[0pt]
W.~Adam, F.~Ambrogi, T.~Bergauer, M.~Dragicevic, J.~Er\"{o}, A.~Escalante~Del~Valle, M.~Flechl, R.~Fr\"{u}hwirth\cmsAuthorMark{1}, M.~Jeitler\cmsAuthorMark{1}, N.~Krammer, I.~Kr\"{a}tschmer, D.~Liko, T.~Madlener, I.~Mikulec, N.~Rad, J.~Schieck\cmsAuthorMark{1}, R.~Sch\"{o}fbeck, M.~Spanring, W.~Waltenberger, C.-E.~Wulz\cmsAuthorMark{1}, M.~Zarucki
% ...
\vskip\cmsinstskip
\dag: Deceased\\
1:  Also at Vienna University of Technology, Vienna, Austria\\
% ...
\end{sloppypar}
\end{document}